# Conditional diffusion-based microstructure reconstruction

Christian Düreth[a,∗], Paul Seibert[b,∗], Dennis Rücker[b], Stephanie Handford[a,c], Markus Kästner[b] and Maik Gude[a,∗∗]

[a]*Institute of Lightweight Engineering and Polymer Technology, TU Dresden, Dresden, 01307, Germany*
[b]*Institute of Solid Mechanics, TU Dresden, Dresden, 01069, Germany*
[c]*University of Mississippi, Oxford, MS 38677, USA*

ABSTRACT

Microstructure reconstruction, a major component of inverse computational materials engineering, is currently advancing at an unprecedented rate. While various training-based and training-free approaches are developed, the majority of contributions are based on generative adversarial networks. In contrast, *diffusion models* constitute a more stable alternative, which have recently become the new state of the art and currently attract much attention. The present work investigates the applicability of diffusion models to the reconstruction of real-world microstructure data. For this purpose, a highly diverse and morphologically complex data set is created by combining and processing databases from the literature, where the reconstruction of realistic micrographs for a given material class demonstrates the ability of the model to capture these features. Furthermore, a fiber composite data set is used to validate the applicability of diffusion models to small data set sizes that can realistically be created by a single lab. The quality and diversity of the reconstructed microstructures is quantified by means of descriptor-based error metrics as well as the Fréchet inception distance (FID) score. Although not present in the training data set, the generated samples are visually indistinguishable from real data to the untrained eye and various error metrics are computed. This demonstrates the utility of diffusion models in microstructure reconstruction and provides a basis for further extensions such as 2D-to-3D reconstruction or application to multiscale modeling and structure-property linkages.

## 1. Introduction

A large number of realistic microstructure data is an important cornerstone to inverse computational materials engineering (ICME) [9]. For this purpose, microstructure characterization and reconstruction, hereafter simply referred to as MCR or reconstruction, is an increasingly active field of research that is focused on (i) generating many microstructures from few examples, (ii) constructing 3D microstructures from 2D slices and (iii) interpolating between microstructures in a morphologically meaningful manner. A wide variety of methods is reviewed in [8, 3, 38] and a brief summary is given in the following. Reconstruction techniques can be divided into more traditional *descriptor-based* approaches and more contemporary *data-based* techniques based on machine learning.

*Descriptor-based* approaches explicitly capture the microstructure morphology by means of so-called microstructure descriptors. Simple volume fractions, Minkowski tensors [42], and spatial correlations [57, 21] are a few examples of these descriptors. After characterizing a microstructure by its descriptors, all other information is discarded and microstructures are reconstructed solely from these descriptors, usually in an iterative optimization procedure. Arguably, the simplest and computationally most efficient strategy consists of randomly placing spherical inclusions into a microstructure until the desired volume fraction is reached [52].

Given further descriptors, these inclusions can be moved to iteratively improve their spatial arrangement. Specifically for fiber reinforced materials, various adapted approaches have been proposed based on an explicitly assumed fiber geometry [30, 41, 29]. More general materials with complex micro-geometries can be approximated by overlapping ellipsoids [40, 56]. In contrast, the Yeong-Torquato algorithm explicitly solves an optimization problem in the space of all possible pixel or voxel values, where the difference between the desired and the current descriptors is minimized by Simulated Annealing [57, 53]. This not only makes the approach conceptually simple and elegant, but also suitable for materials where no explicit inclusion geometry can be defined (e.g. alloys) or where the deviation from the idealized shape is to be considered. However, the high dimensionality of the optimization problem makes it computationally impractical for high-resolution 3D structures. One solution is to limit the descriptors to a computationally cheap subset and develop highly efficient code that performs billions of iterations for materials that can be adequately characterized by the descriptor subset [1]. Alternatively, computationally expensive but differentiable descriptors can be used to reduce the number of iterations by means of gradient-based optimizers. This is described in [43] and [44] as differentiable MCR, and several works can be found that are specific examples of this technique [26, 5, 4]. These methods are implemented in the open-source framework *MCRpy* [45]. Explicitly modeling and harnessing the underlying random field is a less generic but very effective option [14, 27, 37]. Finally, *DREAM.3D* [16] is perhaps by far the most often used descriptor-based MCR tool in practice, despite being

∗Contributed equally
∗∗Corresponding author
maik.gude@tu-dresden.de (M. Gude)
ORCID(s): 0000-0002-6817-1020 (C. Düreth); 0000-0002-8774-8462 (P. Seibert); 0000-0003-3358-1545 (M. Kästner); 0000-0003-1370-064X (M. Gude)

primarily designed for geometric inclusions and grain-like structures.

In *data-based* techniques, a latent representation that is learned from data takes the place of the explicit structural description. After learning this representation in the *training phase*, which is typically expensive, the reconstruction merely requires to query the trained model and can be completed in a matter of seconds or milliseconds. Non-parametric resampling, which was first proposed to the MCR community in [6, 7] and is still being improved upon today [25], is an early example. Recently, however, most attention is directed towards microstructure reconstruction utilizing generative adversarial networks (GANs). Given a training data set, a GAN learns to synthetically generate data points that are not in the training data set, but closely resemble points in the data set [15]. Even on complex image data such as human portraits, remarkable results can be achieved [23]. To the authors' best knowledge, the first application of GANs to microstructure reconstruction dates back to 2017 [31]. In the following, a brief list of exemplary extensions is provided, whereas a separate review paper would be required to cover all extensions and applications. One area of research is enhanced training stability, which is achieved by additional invariances as well as robust and progressively growing architectures [55, 18]. Besides GAN modifications like conditional GANs [46] and BiCycle GANs [13], combinations with other machine learning models are investigated. For example, variational autoencoders can help to smooth the latent space of a GAN [59] and recurrent neural networks in the latent space provide an interesting extension to 3D [58], although it should be mentioned that 2D-to-3D reconstruction can also be achieved by GANs alone [24]. After the success of transformer models in machine learning [54] and their first applications to microstructure reconstruction [60], combinations of GANs and transformers have been shown to extend their range of applicability to higher resolutions [34].

In summary, much attention has been dedicated to GANs and the capabilities regarding scalability and small data sets are continuously increasing. However, some limitations of GANs cannot easily be overcome. Most importantly, GANs often suffer from unstable training due to the adversarial training methods [49, 48]. As an alternative, it is worth noting that much progress is currently made in the machine learning community with *diffusion models*.

Originally inspired by non-equilibrium physics, diffusion probabilistic models were developed by Sohl-Dickstein et al. [47] in 2015, only one year after GANs were developed [15]. As shown in Figure 1, the central idea is to incrementally add small amounts of noise to training images and then learn a reverse diffusion process that can be used to generate structured data from random noise. This avoids the adversarial training methods that are responsible for well-known GAN training issues. Despite this robustness and simplicity, most research on synthetic image generation was focused on GANs, while the foundations for the currently used architectures were developed [39, 10]. Recently, the focus has shifted towards diffusion models after excellent results were achieved in 2020 [20, 48, 49, 22].

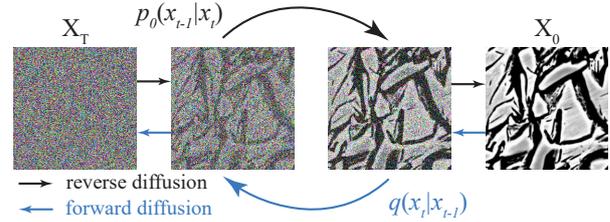

**Figure 1:** Central idea behind diffusion models: A forward diffusion process $q$ incrementally adds noise to an image and a diffusion model is trained to learn the reverse process $p$. The trained model can be repeatedly applied to incrementally generate synthetic data from random noise.

With a few technical improvements regarding the loss function and training procedure, Nichol and Dhariwal at *OpenAI* [32] released an efficient and stable open-source implementation. Validated and benchmarked using well-established image data sets such as CIFAR10 or MNIST, this model and code form a foundation for various applications as well as the present work. For simplicity, the architecture, loss function, and training scheme developed in [32] is hereafter simply referred to as *diffusion model*, while the choice of hyperparameters is discussed later.

In 2021, this code for diffusion models was shown to outperform GANs on common image synthesis tasks [11], which added further momentum to diffusion model research. For example, while the well-known artistic generative model *DALL-E* [36] employs a single-pass transformer-based decoder, the successor, *DALL-E 2* is based on diffusion models because of their efficiency and high-quality samples [35]. Similarly, the recent photorealistic image generation and editing tool *GLIDE* is based on diffusion models [33].

With the foundation of diffusion models proven to be viable, this paper seeks to both, apply them to and validate them on microstructure reconstruction. After presenting the data, model, and hyperparameters in Section 2, the results are presented in Section 3 and summarized in Section 4.

## 2. Methods

The applicability of diffusion models to microstructure reconstruction is validated by two independent data sets described in Section 2.1: While the first data set shows the wide range of applicability of diffusion models to morphologically complex structures, the second data set is much smaller and demonstrates the feasibility for amounts of data that can realistically be generated by a single lab for newly developed materials. After the model is outlined in Section 2.2, the training and choice of hyperparameters are described in Section 2.3.

### 2.1. Data Preparation
#### 2.1.1. Generic Microstructure Data
In order to obtain a large variety of high-quality images, a data set is created by combining the *NFFA-Europe* [2] and the *Ultrahigh Carbon Steel Micrographs* (*UHCS*) [17] data sets. The *NFFA-Europe* data set comprises approximately 21000 images of 10 different classes. The depicted structures range from stochastic media like particles and biological structures or porous sponges to geometrical objects like devices and electrodes that are removed from the data set in this work. The *UHCS* data set contains 961 labeled micrographs of high-carbon steels that were subjected to different heat treatments. Based on the *SQLite* database given in [17], the key *primary_microconstituent* is used to define microstructure classes, such as pearlite and martensite. In summary, six and seven classes are extracted from the data sets, respectively, leading to a total of 13 classes as summarized in Table 1.

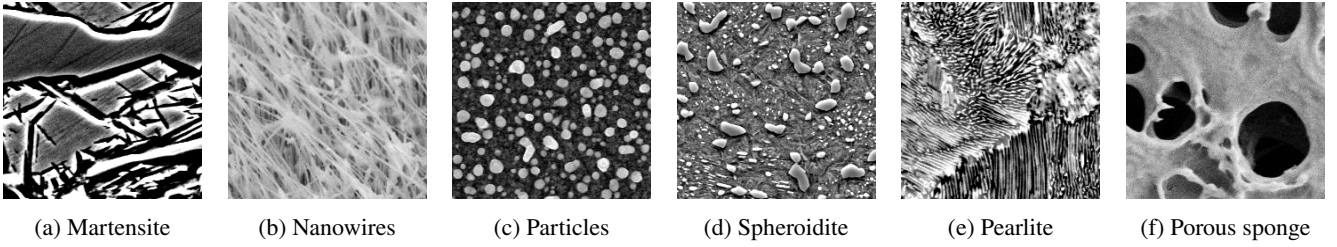

(a) Martensite  (b) Nanowires  (c) Particles  (d) Spheroidite  (e) Pearlite  (f) Porous sponge

**Figure 2:** Example microstructures for six selected classes in the large data set in Section 2.1.1.

**Table 1**
Microstructure classes extracted from different data sets.

| NFFA-Europe [2] | UHCS [17] |
| --- | --- |
| Biological | Martensite |
| Coated surfaces | Pearlite |
| Nanowires | Pearlite + Spheroidite |
| Powder | Pearlite + Widmanstätten |
| Particles | Spheroidite |
| Porous sponge | Spheroidite + Widmanstätten |
|  | Network |

After removing images containing markers or labels, a severe class imbalance is observed, with a factor of over 100 between the largest and smallest class. Consequently, all classes are limited to 1000 samples, and cropping, rotation, and mirroring are used to augment underrepresented classes to reach the same number. The processed data set thus comprises 13000 high-quality images with a resolution of $256 \times 256$ pixels that are equally distributed among 13 classes. Figure 2 shows exemplary members of six selected classes, demonstrating the morphological complexity and diversity of the data.

### 2.1.2. Fiber Composite Data

In order to test the applicability of diffusion models to smaller data sets, a second model is trained based on composite microstructures [12]. For this purpose, a data set is created based on composite microstructure images with a fiber volume content (FVC) of 40%, 50%, and 60%. For each FVC, 36 high-resolution images are available. As schematically shown in Figure 3, 30 quadratic subregions of length and width of $\approx 180\mu m$ pixels are randomly sampled and interpolated to $256 \times 256$ pixels.

### 2.2. Model

The fundamental idea behind diffusion models in image data generation is to *incrementally* apply slight changes to a noisy image until the final result is reached as shown in Figure 1. This stands in contrast to GANs, where a single evaluation of the model suffices to create an image. The present work is based on the Denoising Diffusion Probabilistic Model (DDPM) proposed in [32], which is freely available on *GitHub*. A brief introduction is given in the following, whereas the reader is referred to the relevant sources [47, 20, 32] for further details.

Given a distribution $q(x_0)$ of image data, consider a forward noising process

$$q(x_t|x_{t-1}) = \mathcal{N}\left(x_t; \sqrt{1-\beta_t}x_{t-1}, \beta_t \mathbf{I}\right), \quad (1)$$

which adds Gaussian noise to $x_{t-1}$ from timestep $t-1$ in order to create a slightly noisier version $x_t$ at timestep $t$. Herein, $\mathcal{N}$ denotes the normal distribution, $\beta_t \in (0, 1)$ is the variance and $\mathbf{I}$ is the identity matrix. Repeatedly applying this noising for $T$ timesteps yields

$$q(x_{1..T}) = q(x_0) \prod_{t=1}^{T} q(x_t|x_{t-1}), \quad (2)$$

which can be approximated by a normal distribution if $T$ becomes sufficiently large. Therefore, if the exact reverse distribution $q(x_{t-1}|x_t)$ was known, random noise could be sampled as $x_T \sim \mathcal{N}(0, \mathbf{I})$ and $q(x_{t-1}|x_t)$ could be used to incrementally denoise $x_T$ to obtain a sample from the original distribution $q(x_0)$. The goal of DDPMs is to approximate the reverse distribution by

$$p_\theta(x_{t-1}|x_t) = \mathcal{N}(x_{t-1}; \mu_\theta(x_t, t), \Sigma_\theta(x_t, t)). \quad (3)$$

In particular, Nichol and Dhariwal [32] propose

$$\mu_\theta(x_t, t) = \frac{1}{\sqrt{\alpha_t}}\left(x_t - \frac{\beta_t}{\sqrt{1-\bar{\alpha}_t}}\epsilon_\theta(x_t, t)\right) \quad (4)$$

and

$$\Sigma_\theta(x_t, t) = \exp(v_\theta(x_t, t)\log\beta_t + (1-v_\theta(x_t, t))\log\tilde{\beta}_t), \quad (5)$$

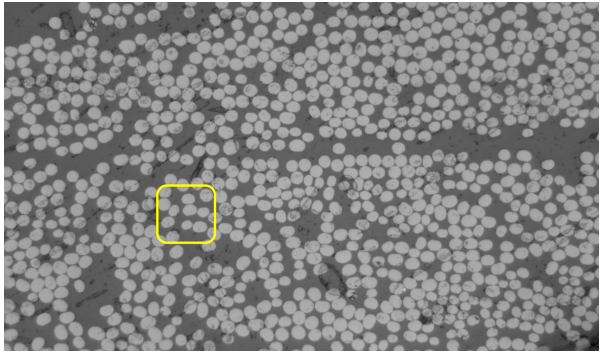

(a) Original sample 40% FVC

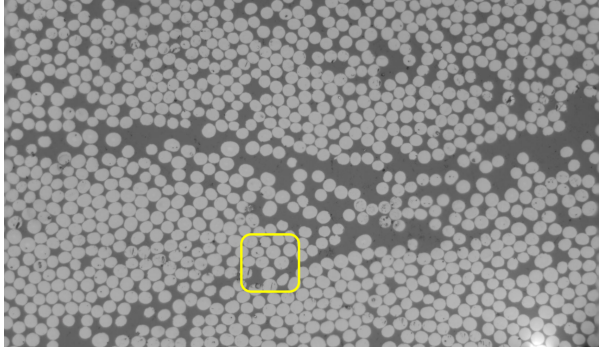

(b) Original sample 50% FVC

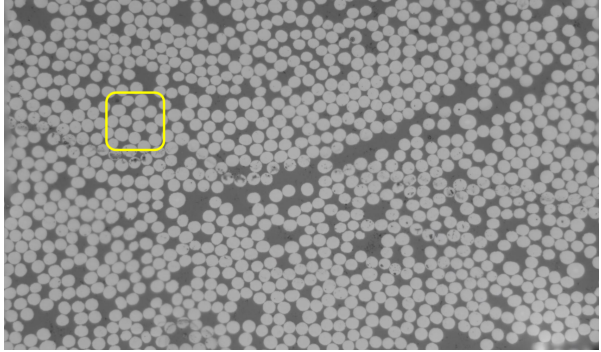

(c) Original sample 60% FVC

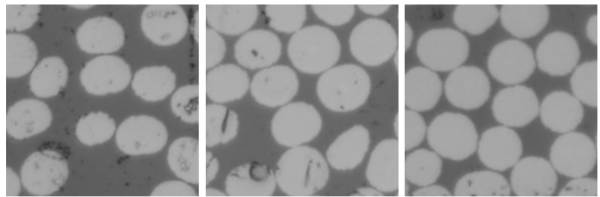

(d) Random subsections used to train the model.

**Figure 3:** Examples of the original fiber data set and some randomly selected subsections.

where $\epsilon_\theta$ and $v_\theta$ are learned by a neural network and the scalar parameters $\alpha_t$, $\overline{\alpha}_t$ and $\tilde{\beta}_t$ are modeled as

$$\alpha_t = 1 - \beta_t \quad , \quad \overline{\alpha}_t = \prod_{s=0}^{t} \alpha_s \quad (6)$$

and

$$\tilde{\beta}_t = \frac{1 - \overline{\alpha}_{t-1}}{1 - \overline{\alpha}_t} \beta_t \ . \quad (7)$$

The motivation for this choice exceeds the scope of this work and is given in the references. Because DDPMs are applied to image data in this work, convolutional neural networks are a natural choice for modeling $\epsilon_\theta$ and $v_\theta$ in Equations (4) and (5), respectively. In particular, to address the need to in- and output high-resolution pixel data, a u-net architecture is chosen using several convolutional residual blocks per resolution level as well as a self-attention block at the 16 × 16 level. Based on the open-source implementation provided by [32], the hyperparameters of the architecture are adapted to microstructure reconstruction as discussed in the following section.

### 2.3. Training and hyperparameters

Like in [32] and [20], the model architecture is fundamentally inspired by *PixelCNN++* [39] and *PixelSNAIL* [10]. To adapt the model to the present task and data set size, a hyperparameter optimization is carried out by comparing models after 80,000 training iterations[1] and selecting the best configuration. The best trade-off between model flexibility and compactness was found at 128 channels and three residual blocks. Maintaining 4000 diffusion steps had the same effect on the data as suggested by [32], a decrease in the log-likelihood. Furthermore, $\Sigma_\theta$ is learned as suggested in [32]. In contrast to that work, the linear noise schedule is observed to slightly outperform the cosine schedule, hence, the linear noise scheduler is chosen. Finally, class conditioning is enabled for a classifier-free guidance, hence, the model is conditioned on the desired class alone, as opposed to classical, descriptor-based reconstruction methods, where a high-dimensional descriptor vector is given. While more complex conditioning is also possible for diffusion models [33], we leave this extension for microstructure reconstruction open for future work. The final set of hyperparameters is summarized in Table 2 and the corresponding model is trained for a total of 300,000 iterations.

In total, two independent models are fully trained, model I on the diverse data set of Section 2.1.1 and model II on the smaller fiber data set of Section 2.1.2. Both models have the same hyperparameters, although the hyperparameter optimization is carried out only on the large data set. Furthermore, both models are trained from scratch, although transfer learning might be an interesting approach for future investigations regarding small data set sizes. Fully training the final model takes approximately 48 hours on a single *Nvidia A100* GPU with a batch size of 8 to make full use of the 40 GB memory. On the same hardware, reconstructing 64 microstructures per class with a resolution of 256 × 256 pixels from the trained model takes 16 hours. This amounts to approximately 70 seconds per microstructure, although it should be noted that a batch size of 64 is used to utilize the available memory.

---
[1]This allows for a computationally efficient hyperparameter optimization while still being representative for the final model performance.

**Table 2**
Parameters of diffusion model.

| Parameter | Value |
| --- | --- |
| num_channels | 128 |
| num_res_blocks | 3 |
| diffusion_steps | 4000 |
| learn_sigma | True |
| noise_schedule | Linear |
| class_cond | True |

## 3. Results

Table 3 shows one generated structure per class in comparison with similar micrographs in the training data set. The synthetic structures are indistinguishable from real data to the untrained eye and the wide range of complex morphologies from porous sponges to martensite and particles is captured as well as the real deviation of the fiber cross-sections from their idealized circular shape. However, it is clear that objective metrics are required to evaluate the results beyond visual impressions. Therefore, the generated microstructures are analyzed in terms of sample *quality* and *diversity*.

To quantify the image *quality*, i.e. the similarity of the reconstructed microstructures to images in the training data set, descriptor errors are defined in terms of the spatial three-point correlation [21] $S$

$$\mathcal{E}_S = \min_{M_i \in \mathcal{M}^{\text{train}}} ||S(M_i) - S(M^{\text{rec}})||_{\text{MSE}} \quad (8)$$

and Gram matrices [28] $G$

$$\mathcal{E}_G = \min_{M_i \in \mathcal{M}^{\text{train}}} ||G(M_i) - G(M^{\text{rec}})||_{\text{MSE}}, \quad (9)$$

whereby distances are defined by the mean squared error (MSE) and the closest microstructure example $M_i$ in the training data set $\mathcal{M}^{\text{train}}$ is chosen. The open-source software *MCRpy* [45] is used to compute these descriptors as

```
settings = mcrpy.CharacterizationSettings(
    descriptor_types=['Correlations', 'GramMatrices'],
    use_multigrid_descriptor=False, limit_to=32,
    use_multiphase=False)
descriptors = mcrpy.characterize(microstructure, settings)
```

To demonstrate that the images are not present in the training data set, an additional error metric

$$\mathcal{E}_{px} = \min_{M_i \in \mathcal{M}^{\text{train}}} ||M_i - M^{\text{rec}}||_{\text{MSE}} \quad (10)$$

is defined directly in pixel space. For each material class (13 classes for model I and 3 classes for model II), these three error metrics are computed for all 64 generated samples. The median of all generated samples as well as the specific error of the examples in Table 3 are given in Table 4. Furthermore, the micrographs in the training set that minimize $\mathcal{E}_S$ and $\mathcal{E}_G$ for the selected generated sample in Table 3 are also given for reference. Table 4 shows that the errors are very small although, unlike with classical, descriptor-based reconstruction methods [57, 43], they are not directly prescribed to the model or in any other way incorporated in the training. Furthermore, it can be seen that the generated micrographs are not present in the training data set, but are genuinely independent samples. This follows from $\mathcal{E}_S > 0$ and $\mathcal{E}_G > 0$ but becomes even more obvious by $\mathcal{E}_{px} > 0$, which is the main reason to list this metric. Finally, it becomes clear that the example micrographs displayed in Table 3 are not "cherry-picked" as the best results, since the example's error is sometimes above the median and sometimes below. Nevertheless, it should be mentioned that the median is given in Table 4 instead of the mean of the distribution because some of the generated structures have errors that are orders of magnitude larger. These outliers are not problematic, however, since the trained diffusion model is relatively inexpensive to evaluate and insufficient results can be simply discarded. Finally, it should also be explained why some of the training data set examples in Table 3 that minimize $\mathcal{E}_S$ are visually significantly less similar to the corresponding generated sample than the example that minimizes $\mathcal{E}_G$. This occurs due to the fact that unsegmented micrographs are used for the computation of the descriptor and the spatial $n$-point correlations are significantly more sensitive to the volume fraction, i.e. the image brightness, than the Gram matrices. In summary, the evaluation of the descriptor errors allows us to conclude that a very high *quality* is achieved throughout all classes, both, for the complex morphologies trained in model I as well as the relatively small fiber data set used for model II.

To quantify the image *diversity*, the Fréchet inception distance (FID) score is chosen as an established error metric from the machine learning community [19]. In contrast to the previous errors, which compare individual microstructures, the FID score quantifies the similarity of data distributions[2]. Specifically, the mean values $\mu_i$ and covariances $C_i$ of the internal activations of the deepest layer of the *Inceptionv3* network [51] are used to compute the difference between two data sets ($i = 1, 2$) as

$$\text{FID} = ||\mu_1 - \mu_2||_2^2 + \text{Tr}\left(C_1 + C_2 - 2\sqrt{C_1 \cdot C_2}\right) \quad (11)$$

where Tr refers to the trace. Table 4 shows the FID scores of both models for each class, which approximately range between 100 and 300 for model I and between 50 and 100 for model II. In contrast, the same diffusion model achieved an FID score of 30 on ImageNet [32]. Hence, the reported FID score is not on par with state-of-the-art machine learning models for general image generation but suffices to conclude that the generated samples are indeed diverse. Furthermore, it needs to be kept in mind that the data sets used in this work are relatively small by machine learning standards and that the FID score is computed based on the internal activations of a network that was never trained on micrograph data. Nevertheless, in order to substantiate the claim of sample diversity, 11 additional samples per class are shown in Table 5. In summary, the results are convincing in both, image quality and diversity.

---

[2]N.B.: Hence, this metric does not quantify diversity alone, but a combination of image quality and diversity. A suitable and well-established way to quantify diversity independently from quality is not known to the authors.

**Table 3**
An exemplary generated sample per material class. Note the distinction between model I, which is trained on the diverse data set from Section 2.1.1 and model II, which is trained on the fiber data set from Section 2.1.2. For each generated sample, examples from the training data set are shown that are closest in terms of the spatial correlations $S$ and the Gram matrices $G$, respectively.

| | Model I | | | | | | | |
|---|---|---|---|---|---|---|---|---|
| | Biological | Coated surfaces | Nanowires | Powder | Particles | Porous sponge | Martensite | Pearlite |
| generated | 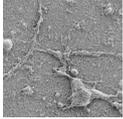 | 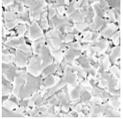 | 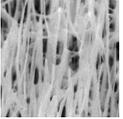 | 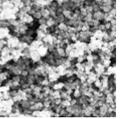 | 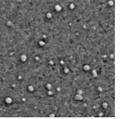 | 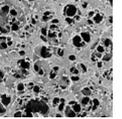 | 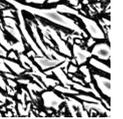 | 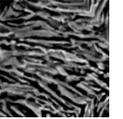 |
| closest in $S$ | 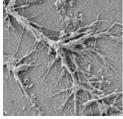 | 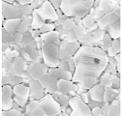 | 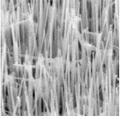 | 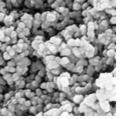 | 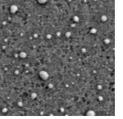 | 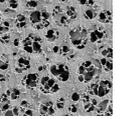 | 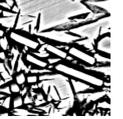 | 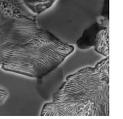 |
| closest in $G$ | 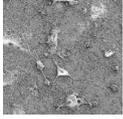 | 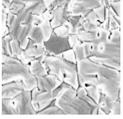 | 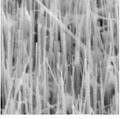 | 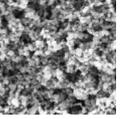 | 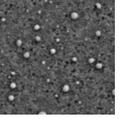 | 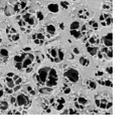 | 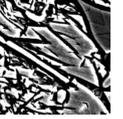 | 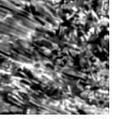 |

| | Model I | | | | | Model II | | |
|---|---|---|---|---|---|---|---|---|
| | Pearlite + Spheroidite | Pearlite + Widmanstätten | Spheroidite | Spheroidite + Widmanstätten | Network | 40% FVC | 50% FVC | 60% FVC |
| generated | 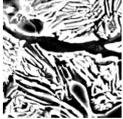 | 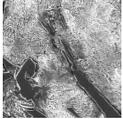 | 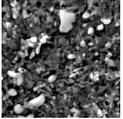 | 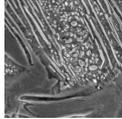 | 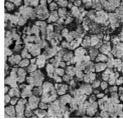 | 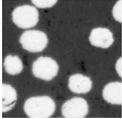 | 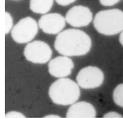 | 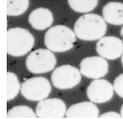 |
| closest in $S$ | 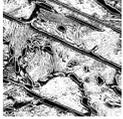 | 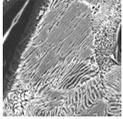 | 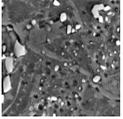 | 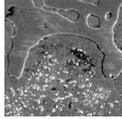 | 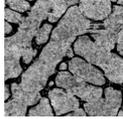 | 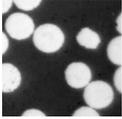 | 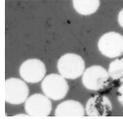 | 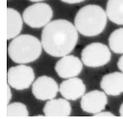 |
| closest in $G$ | 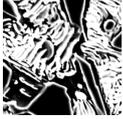 | 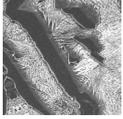 | 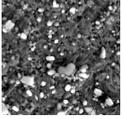 | 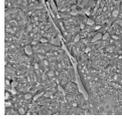 | 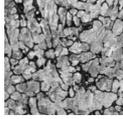 | 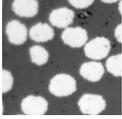 | 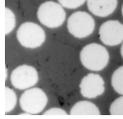 | 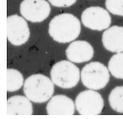 |

## 4. Conclusion

The applicability of diffusion models to microstructure reconstruction is demonstrated based on two real-world micrograph data sets. As a *first* step, a very diverse and complex data set is created by combining and processing databases from the literature, with 13 classes ranging from biological, porous sponge, and particles to martensite, pearlite, and spheroidite. As a *second* step, a fiber composite data set is used to validate the applicability of diffusion models to small data sets that can realistically be created by a single lab. The quality and diversity of the reconstructed microstructures are quantified by means of descriptor-based error metrics as well as the FID score. Although not present in the training data set, the generated samples are visually indistinguishable from real data to the untrained eye and the descriptor errors are very small. While the FID score is not on par with state-of-the-art machine learning models for general image synthesis, it has to be kept in mind that comparably small data sets are used in the present work and that the applicability of FID scores to the given data is not clear. In summary, diffusion models can be considered to be capable of capturing complex microstructural features and reconstructing structures from random noise.

**Table 4**
Descriptor errors and FID score for each material class. For the descriptor errors, the value of the specific example in Table 3 is shown as well as the median of all 64 generated micrographs (in parentheses). Lower is better.

| | Class | $\mathcal{E}_S$ in % Example | Median | $\mathcal{E}_G$ in % Example | Median | $\mathcal{E}_{px}$ in % Example | Median | FID |
|---|---|---|---|---|---|---|---|---|
| Model I | Biological | $4 \cdot 10^{-5}$ | $(5 \cdot 10^{-5})$ | 0.5 | (0.5) | 1.2 | (1.4) | 181 |
| | Coated surfaces | $9 \cdot 10^{-2}$ | $(2 \cdot 10^{-4})$ | 1.7 | (0.7) | 1.8 | (1.1) | 144 |
| | Nanowires | $9 \cdot 10^{-4}$ | $(9 \cdot 10^{-4})$ | 2.8 | (4.6) | 2.6 | (4.1) | 188 |
| | Powder | $6 \cdot 10^{-4}$ | $(6 \cdot 10^{-4})$ | 0.7 | (1.0) | 5.3 | (4.8) | 107 |
| | Particles | $6 \cdot 10^{-6}$ | $(6 \cdot 10^{-6})$ | 0.3 | (0.9) | 1.0 | (1.9) | 203 |
| | Porous sponge | $7 \cdot 10^{-5}$ | $(9 \cdot 10^{-5})$ | 4.6 | (4.1) | 6.8 | (5.4) | 172 |
| | Martensite | $1 \cdot 10^{-2}$ | $(2 \cdot 10^{-3})$ | 12 | (5.9) | 17 | (7.6) | 203 |
| | Pearlite | $2 \cdot 10^{-5}$ | $(4 \cdot 10^{-4})$ | 11 | (4.3) | 5.7 | (5.3) | 197 |
| | Pearlite + Spheroidite | $9 \cdot 10^{-3}$ | $(2 \cdot 10^{-3})$ | 19 | (6.2) | 16 | (7.3) | 216 |
| | Pearlite + Widmanstätten | $2 \cdot 10^{-3}$ | $(2 \cdot 10^{-4})$ | 2.3 | (2.7) | 5.1 | (4.2) | 296 |
| | Spheroidite | $2 \cdot 10^{-4}$ | $(3 \cdot 10^{-5})$ | 3.0 | (1.4) | 5.6 | (2.8) | 205 |
| | Spheroidite + Widmanstätten | $5 \cdot 10^{-5}$ | $(2 \cdot 10^{-4})$ | 3.4 | (2.7) | 3.0 | (4.2) | 231 |
| | Network | $4 \cdot 10^{-5}$ | $(4 \cdot 10^{-5})$ | 0.7 | (1.3) | 3.9 | (2.3) | 106 |
| Model II | 40% FVC | $2 \cdot 10^{-3}$ | $(2 \cdot 10^{-2})$ | 5.3 | (3.7) | 1.2 | (1.8) | 57 |
| | 50% FVC | $3 \cdot 10^{-3}$ | $(3 \cdot 10^{-2})$ | 2.7 | (2.9) | 1.9 | (1.7) | 89 |
| | 60% FVC | $3 \cdot 10^{-3}$ | $(3 \cdot 10^{-2})$ | 4.8 | (3.8) | 1.7 | (1.8) | 103 |

In this context, it has to be stressed that training stability is a major advantage of diffusion models over generative adversarial networks, which are currently very common in data-based microstructure reconstruction: The authors have not experienced a single incidence of training instability during hyperparameter optimization.

These findings motivate further work such as the extension to 2D-to-3D reconstruction or to even smaller data set sizes. Furthermore, the application to segmented data is a relevant topic. In this context, in view of the success of the underlying u-net model in segmenting microstructural data [50], a hybrid method to perform simultaneous reconstruction and segmentation could be conceivable. Finally, the extension of the simple class conditioning in the present work to advanced conditioning based on real-valued higher-dimensional and general microstructure descriptors will pave the way to numerous applications in multiscale modeling, exploration of structure-property linkages, and computational materials engineering.

## Acknowledgements

The groups of M. Gude and M. Kästner thank the German Research Foundation DFG which supported this work under Grant numbers GU 614/35-1 | KA 3309/17-1 and GU 614/30-1. All presented computations were performed on a HPC-Cluster at the Center for Information Services and High Performance Computing (ZIH) at TU Dresden. The authors thus thank the ZIH for generous allocations of computer time.

## CRediT authorship contribution statement

**Christian Düreth:** Conceptualization, Data curation, Investigation, Methodology, Software, Supervision, Visualization, Writing - review and editing. **Paul Seibert:** Conceptualization, Formal analysis, Methodology, Supervision, Validation, Visualization, Writing - original draft preparation, Writing - review and editing. **Dennis Rücker:** Data curation. **Stephanie Handford:** Investigation, Visualization. **Markus Kästner:** Funding acquisition, Resources, Writing - review and editing. **Maik Gude:** Funding acquisition, Resources, Writing - review and editing.

**Table 5**
Multiple generated samples per class show the diversity of the trained diffusion models.

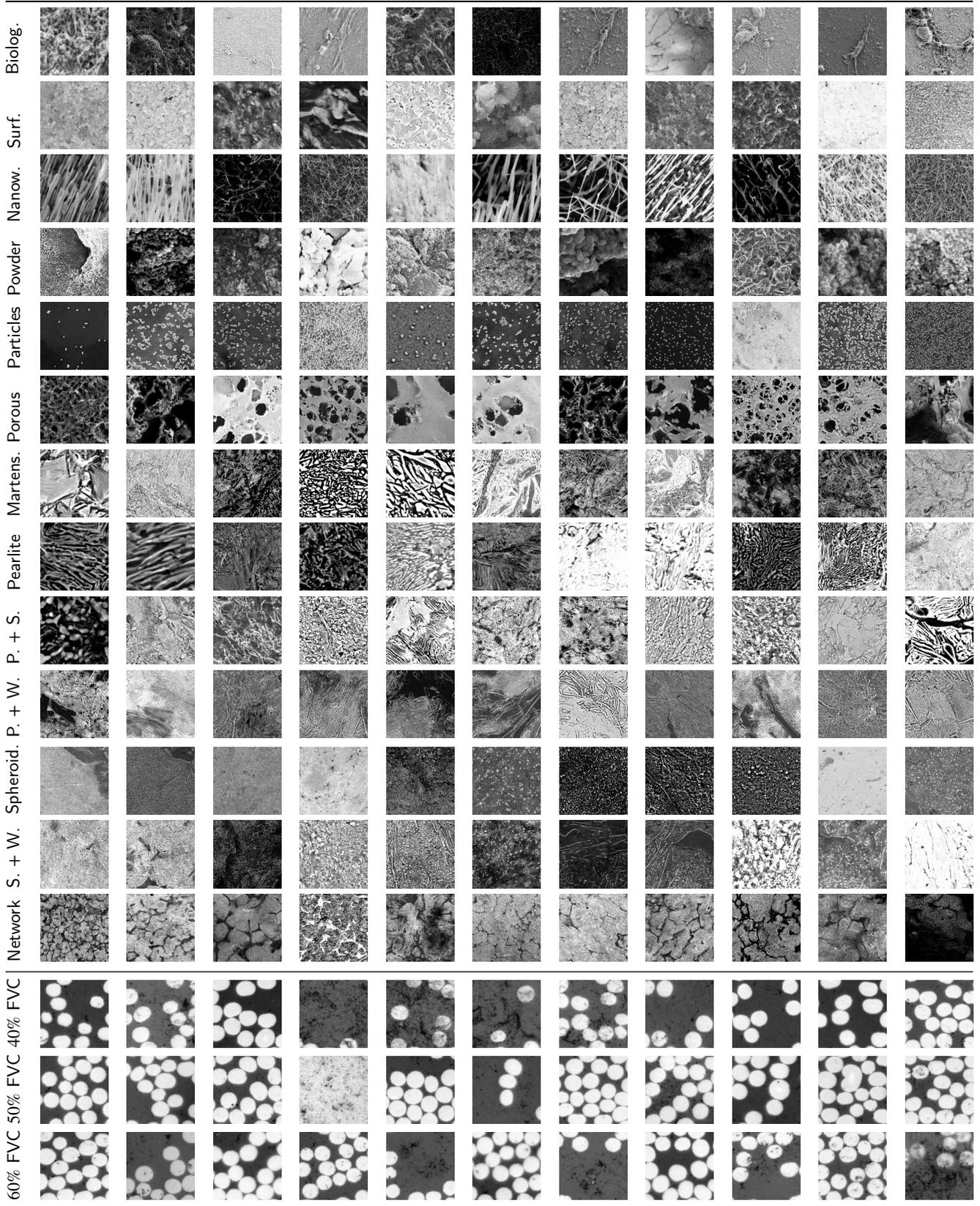